\documentclass[aps,prb,twocolumn,superscriptaddress]{revtex4-1}

\usepackage{graphicx}
\usepackage{xcolor}
\usepackage{dcolumn}
\usepackage{bm}
\usepackage{amsmath}
\usepackage{amssymb}

\begin{document}

\title{Cycling and spiral-wave modes in an active cyclic Potts model}

\author{Hiroshi Noguchi}
\email[]{noguchi@issp.u-tokyo.ac.jp}
\affiliation{Institute for Solid State Physics, University of Tokyo, Kashiwa, Chiba 277-8581, Japan\\
corresponding author: noguchi@issp.u-tokyo.ac.jp}
\author{Fr\'ed\'eric van Wijland}
\email[]{frederic.van-wijland@u-paris.fr}
\affiliation{Laboratoire Mati\`ere et Syst\`emes Complexes (MSC), Universit\'e Paris Cit\'e \& CNRS, 75013 Paris, France}
\author{Jean-Baptiste Fournier}
\email[]{jean-baptiste.fournier@u-paris.fr}
\affiliation{Laboratoire Mati\`ere et Syst\`emes Complexes (MSC), Universit\'e Paris Cit\'e \& CNRS, 75013 Paris, France}

\begin{abstract}
We studied the nonequilibrium dynamics of a cycling three-state Potts model using simulations and theory. This model can be tuned from thermal-equilibrium to far-from-equilibrium conditions. At low cycling energy, the homogeneous dominant state cycles via nucleation and growth, while spiral waves are formed at high energy. For large systems, a discontinuous transition occurs from these cyclic homogeneous phases to spiral waves, while the opposite transition is absent. Conversely, these two modes can coexist for small systems. The waves can be reproduced by a continuum theory, and the transition can be understood from the competition between nucleation and growth.
\end{abstract}

\maketitle

\section{Introduction}

In two-dimensional systems, target or spiral waves can be observed in specific nonequilibrium conditions.\cite{mikh94,kikh96,rabi00,murr03,mikh06,okuz03,qu06,sugi15} One of the conditions for these to emerge is the existence of several states (three or more) along with the possibility of cyclic transitions between these. This cyclic requirement is a physical reality in various systems ranging from chemical reactions, gene transcription, and food chains.\cite{mikh06,mikh09,nova08,kerr02} These phenomena are well-captured by a description in terms of nonlinear but deterministic partial differential equations. Spatiotemporal patterns under the influence of noise and fluctuations have been much less explored. Noise is able to trigger excitable waves or to generate stochastic resonance between macroscopic states.\cite{kikh96,gamm98,mcdo09,hohm96} Similarly, wave propagation of subexcitable media of photosensitive Belousov--Zhabotinsky reaction is enhanced by the random variations of light intensity.\cite{mikh06,kada98,alon01} 
Noises can be included in lattice-based simulations, cell automata. They have been applied to 
predator--prey systems (also called lattice Lotka--Volterra model and rock--paper--scissors game).\cite{kerr02,szol14,itoh94,tain94,szab99,szab02,reic07,reic08,szcz13,kels15,mir22}
Spatial and time correlations\cite{itoh94,reic08} and vortex dynamics\cite{szab99,szab02} have been investigated.
However, the pair update process in these models is not accompanied by a reverse process (there is no local detailed balance), so that they are not applicable to thermal equilibrium and near-equilibrium conditions. 
In small systems, where thermal fluctuations play an important role, nucleation and growth drive the kinetics of the whole system. However, how such kinetic and stochastic building blocks compete with spatiotemporal patterns is an unchartered territory (see Refs.~\citenum{mikh06,zwkc22} for an introductory discussion). Our purpose in this work is to investigate how thermal noise can drive transitions between dynamical states in a background allowing for spatiotemporal patterns. Our model can be tuned arbitrarily far from or close to equilibrium. We find that spiral waves can be subsumed by noise into the cyclic nucleation and growth of domains.

The model and methods are described in Sec.~\ref{sec:model}.
Simulation results  are presented and discussed in Sec.~\ref{sec:results}.
A continuum theory is described  in Sec.~\ref{sec:theory}.
The relation with the predator--prey models is discussed in Sec.~\ref{sec:predator}.
Finally, a summary is presented in Sec.~\ref{sec:sum}.

\section{Cyclic Potts model}\label{sec:model}

\begin{figure}[tbh]
\includegraphics[]{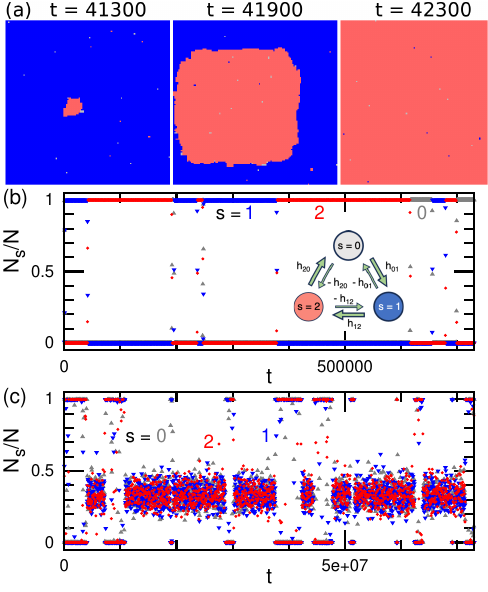}
\caption{
Time evolution of the cyclic Potts model for $h=0.9$ (a)--(b), and $h=1.1$ (c) at $N=128^2$.
(a) Sequential snapshots of the nucleation and growth of an $s=2$ (red [medium gray]) state at short times in a sea of $s=1$ (blue [dark gray]).
(b) Cyclic changes of the dominant states (short span). The fraction of $s$ spins $N_s/N$ is shown as a function of time.  The initial state is blue with $N_1/N\sim O(1)$, while $N_0/N$ ($s=0$ is gray) and $N_{2}/N$  (red) are almost zero; the transitions between different states are very short.
(c) At slightly higher $h$, after an initial HC behavior, the system exhibits stochastic back-and-forth changes between the HC and the SW modes (cloud of colored points). 
{A schematic of the cyclic dynamics of the three states is pictured in the inset of (b).}
}
\label{fig:t09}
\end{figure}

To flesh our ideas out, we consider a two-dimensional square lattice with sites $i$ that can be found in either of three states ($s_i=0, 1$ and $2$).
Each site has four nearest neighbors.
The neighboring sites have a contact energy $-J$ if they are in the same state and $0$ 
if they are in different states.
In addition, the three states have different self-energies $\varepsilon_0, \varepsilon_1$ 
and $\varepsilon_2$, so that the total Hamiltonian reads
\begin{equation}
H = H_{\mathrm{int}} + \sum_i\varepsilon_{s_i},\;\;H_{\mathrm{int}}= -J\sum_{\langle ij\rangle} \delta_{s_i,s_j}.
\end{equation}
This describes a Potts model with external fields.\cite{pott52,bind80}
We define the flipping energies $h_{ss'}$ as the variations $h_{ss'}=\varepsilon_{s}-\varepsilon_{s'}$. 
Hence, by definition $h_{01}+h_{12}+h_{20}=0$. 
\begin{figure}
\includegraphics[]{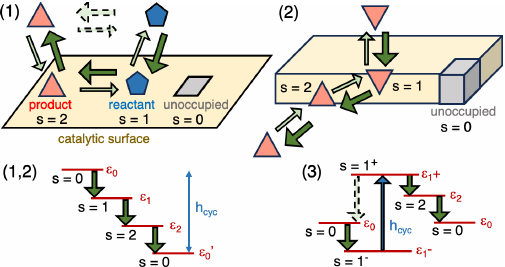}
\caption{
Schematic of the systems. The upper panels show (1) a reaction on a catalytic surface and (2) a molecular transport through a membrane.
The left and right lower panels show the energy levels in the case of (1) and (2), and in the case of (3) involving an excitation, respectively.
}
\label{fig:cart}
\end{figure}
We have several physical realizations in sight where the nonequilibrium drive operates in much the same way as in our Potts model. They are shown in Fig.~\ref{fig:cart}. Consider for instance molecules binding and unbinding to a surface, with the state $s=0$ corresponding to an empty surface site, and the other two states to a site occupied by a molecule, either in the form $s=1$ or $s=2$. A first possibility for dynamical evolution is that the surface catalyzes the reaction from $s=1$ to $s=2$, while in the bulk this reaction is kinetically frozen (case 1 in Fig.~\ref{fig:cart}). This brings the system out of equilibrium.
A second realization can be achieved with proteins binding to the two sides of a bilayer membrane (states $s=1,2$) and switching between them (flip--flop). A nonequilibrium drive is achieved by imposing a chemical potential difference between the solutions on {upper and lower} sides of the membrane {(case 2 in Fig.~\ref{fig:cart})}.\cite{miel20,holl21,nogu23} A third realization, involving an active process, is that $s=1$  has a low-energy state $s=1^-$ and a high-energy state $s=1^+$ triggered by external means (such as photoexcitation or ATP hydrolysis). Then if the reverse transformation from $s=1^+$ to $s=0$ is negligibly slow, we effectively drive the system out of equilibrium (case 3 in Fig.~\ref{fig:cart}). 
Experimentally, chemical waves have been observed at catalytic surfaces (H$_2$ or CO oxidation and NO or NO$_2$ reduction on noble metal surfaces).\cite{ertl08,bar94,barr20,zein22}
Although more than two reactions typically occur on the surfaces, our model corresponds to a single reaction. 
As a final example, a target wave was induced by the transfer of water molecules through a chiral liquid-crystalline monolayer.\cite{tabe03}
The common thread to these physical realizations is that the dynamics of the underlying Potts model is endowed with nonequilibrium features. In practice, we adopt transition rates $w(s_i\to s_i')$ for the states that have a local equilibrium form involving the energy difference  
\begin{equation}\label{eq:defDeltaH}\Delta H_{s_is'_i}= H'_{\mathrm {int}}-H_{\mathrm{int}}-h_{s_is'_i}
\end{equation}
namely  ${w(s_i\to s_i')}/{w(s_i'\to s_i)}=\exp(-\Delta H_{s_is'_i})$ ({\it e.g.} in a Metropolis or in a Glauber rate), taking $h_{s_is'_i}=-h_{s'_is_i}$ but $h_{01}+h_{12}+h_{20}=h_{\mathrm{cyc}}\neq 0$, which acts as the driving field. Locally, the detailed-balance condition, $h_{s'_is_i}=-h_{s_is'_i}$ is maintained, as illustrated in the inset of Fig.~\ref{fig:t09}(b). Coming back to our physical examples, the role of $h_{\mathrm{cyc}}$ is played by the difference of the energies of the two species in the bulk in our first example; in our second set, this is a difference in the chemical potentials between two sides of the membrane; in the third one, this is the energy injected by the active process at work.

For simplicity, we restrict the present analysis to a simple condition with cyclic symmetry, {\it i.e.} $h_{01}= h_{12}= h_{20} = h$. The thermal energy and the lattice spacing are normalized to unity.
We use $J= 2$ to induce a phase separation between different states.
Our lattice has $N$ sites in a square box with periodic boundary conditions.
The state of a variable is flipped according to a Monte Carlo (MC) method but no spin-exchange (diffusion) is considered.
A site is chosen at random, and then it is flipped to either of the other two states with $1/2$ probability.
The new state is accepted or rejected with Metropolis probability
\begin{equation}\label{eq:mc}
p_{s_is_i'}=\min\left(1,e^{-\Delta H_{s_is'_i}}\right)
\end{equation}
This procedure is performed $N$ times per MC step (time unit). Statistical errors are calculated from three or more independent runs.

\section{Simulation Results}\label{sec:results}

When $h=0$, the system is an equilibrium Potts model at low temperature, with three equivalent phases $s=0,1,2$. For brevity, let us set $[k+1]=(k+1)\mod3$. When $h\ne0$, since $h_{k,[k+1]}>0$ and $h_{[k+1],k}=-h_{k,[k+1]}$, the system will exhibit cyclic transformations, the macroscopic state $s=k$ transforming to $s=[k+1]$. However, in a uniform $s=k$ phase, a flip of a single site to $s=[k+1]$ will not easily occur, because of the interaction energy that favors like-states. A cooperative nucleation event, involving the flipping of several sites, is needed to overcome the created boundary energy. The system thus harbors a competition between nucleation and nonequilibrium cycling. Based on these premises, we expect two dynamical modes. (i) A homogeneous cycling mode (HC) in which transient domains of state $s=[k+1]$ nucleate and grow inside domains of state $s=k$. In near-thermal-equilibrium conditions, this will lead to cycling uniform states. (ii) A spiral wave (SW) mode. The latter should take place when a triple point appears, {\it i.e.}  when three phases and three boundary lines meet at the same location in space. A spiral wave should then be generated due to the cyclic $s=k$ to $s=[k+1]$ transition, without the need for nucleation, since boundaries are present. With these scenarios in mind, we now move to a quantitative exploration upon varying the amplitude of $h$ and the system size.

At low $h$, for medium-size systems, starting from a homogeneous initial state, standard domain growth is put off-balance by the cyclic condition. In practice, once a given state wins over, owing to a long nucleation time (see Fig.~\ref{fig:t09}(a) and Movie S1), no domain with a different order can emerge within it. This uniform dominant state cyclically changes ($s=0 \to 1 \to 2 \to 0\ldots$), as  shown in Fig.~\ref{fig:t09}(b). For most of the time, one state is dominant, but  nucleation and growth stochastically occur. This new cyclic phase change, HC mode, is intrinsically noise-driven and does not occur in deterministic systems nor in the cyclic predator--prey models.
In the predator--prey models, extinction corresponds to an absorbing state.
In deterministic systems, periodic oscillations are obtained. 
Adding noise disturbs the oscillations, as observed in a continuous flow stirred tank reactor (CSTR),\cite{hohm96}
but in a qualitatively different way from the cyclic change in the HC mode.

\begin{figure}[tbh]
\includegraphics[]{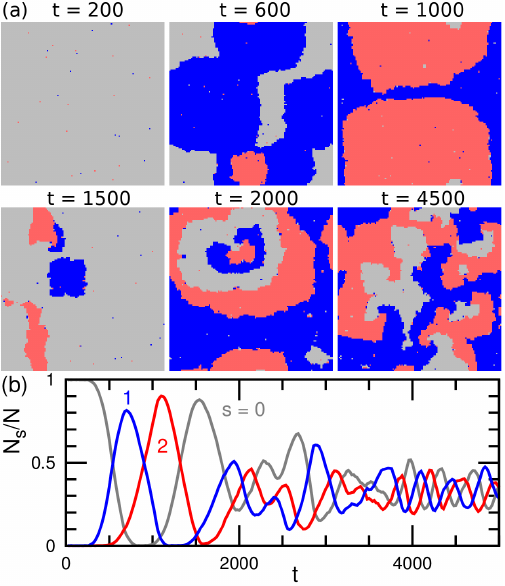}
\caption{
Formation of spiral waves at $h=1.5$ and  $N=128^2$.
(a) Sequential snapshots. Light gray, blue [dark gray], and red [medium gray]
represent $s=0$, $1$ and $2$, respectively.  {The first three snapshots show transient dynamics from a homogeneous phase ($s=0$). } The birth of the SW phase can be spotted in the $t=1500$ snapshot with the existence of a triple point, which eventually leads to a mature SW phase.
(b) Time evolution of the ratios of the three states in the simulation (a), showing {the formation of} the SW mode at $t\simeq2000$.
}
\label{fig:t15}
\end{figure}

\begin{figure}[tbh]
\includegraphics[]{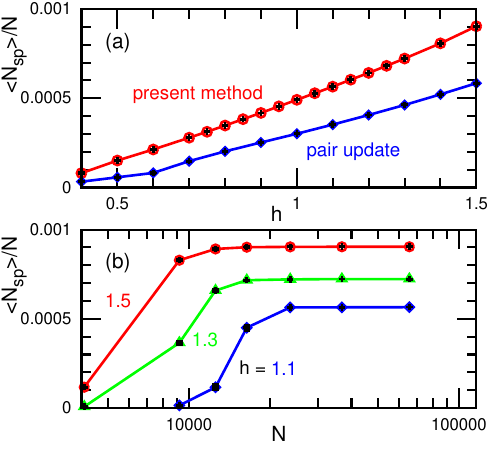}
\caption{
Density $N_{\mathrm{sp}}/N$ of the centers of spiral waves.
(a) Dependence on $h$ at $N=256^2$.
(b) Dependence on $N$ at $h=1.1$, $1.3$, and $1.5$.
The red circles in (a) and all symbols in (b) represent the density obtained by the present method (single update process).
The blue diamonds in (a) represent the density obtained by the pair update process used in the predator--prey models.
}
\label{fig:nsp}
\end{figure}

As $h$ increases, the system exhibits stochastic transitions between the HC mode and the SW mode, as depicted in Fig.~\ref{fig:t09}(c). 
At high $h$, the HC mode is unstable and the SW mode soon appears,  even from a uniform dominant state as shown in Fig.~\ref{fig:t15} and Movie S2.
A circular $s=1$ domain forms by nucleation and spreads to the whole lattice; in the meantime, an  $s=2$ domain appears and grows inside an $s=1$ domain. Subsequently, a spiral wave initiates from the contact of three states.
Clockwise and anticlockwise spirals are generated, as observed in the predator--prey systems\cite{szab99} (see the left-bottom panel of Fig.~\ref{fig:t15}(a)).
The wavelength is broadly distributed owing to the stochastic nature of the fundamental processes.
When a diphasic $s=1,2$ band is set as an initial condition in a $s=0$ background, it first travels as a one-dimensional wave (see the theory below), then eventually transforms into spiral waves by interacting with nucleating domains (see Movie S3). It is because at the contact of three states, the three meeting bands travel circularly that they build a spiral wave.
In the SW mode, all states saturate to a fraction around $1/3$ (see Figs.~\ref{fig:t09}(c) and \ref{fig:t15}(b)).
The density $N_{\mathrm{sp}}/N$ of the three-contact points (i.e., the vortices or centers of the spiral waves) increases with increasing $h$
but is constant for different sizes in the SW mode (see Fig.~\ref{fig:nsp}).
To transform into the HC mode, the three-contact points are required to disappear.
The stochastic disappearance of the spiral centers results in the spreading of one domain (see Movie S4).
As expected from the snapshots of  Fig.~\ref{fig:t15}(a), this scenario occurs frequently at $N=64^2$, 
whereas it was not observed  at $N=256^2$.
And indeed, the lifetime $\tau_{\mathrm{wave}}$ of spiral waves depends exponentially on the system size $N$ (see Fig.~\ref{fig:twave} and text in supplementary information).
Similar exponential-size dependence was reported for the lifetime of transient spiral chaos in excitable media.\cite{qu06,sugi15}
Although this annihilation is an absorbing transition in the excitable media and predator--prey systems,\cite{szol14,reic07} in our system the SW mode is not absorbing and can transform back into the HC phase via nucleation and growth. 
Note also that target patterns can appear transiently during the transition from the SW to HC modes.

\begin{figure}[tbh]
\includegraphics[]{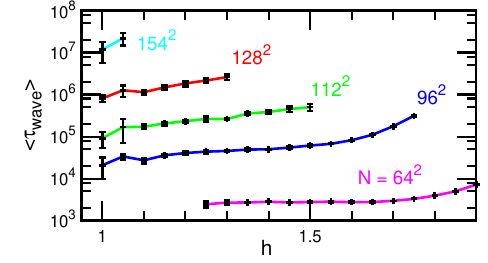}
\caption{
Mean duration $\langle\tau_{\mathrm{wave}}\rangle$ of spiral waves for $N=64^2$, $96^2$,  $112^2$, $128^2$, and $154^2$.
}
\label{fig:twave}
\end{figure}

\begin{figure}[tbh]
\includegraphics[]{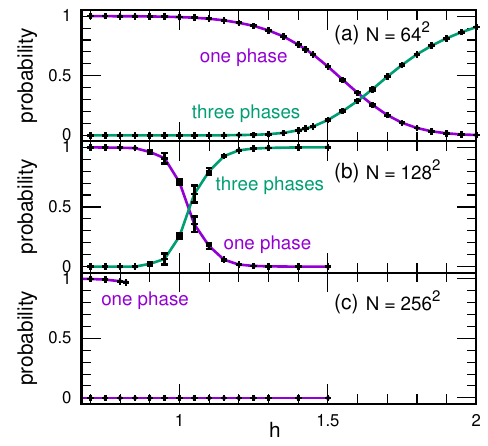}
\caption{For a given $h$, we measure the fraction of the time  spent in one homogeneous phase (defined by the fact that for at least one $s$ there is $N_s>0.98N$) and the fraction  of time where three phases coexist (SW mode: $N_0>0.02N$, $N_1>0.02N$, and $N_2>0.02N$). These occupation probabilities are shown in violet and green, respectively, as a function of  $h$. As the system size is increased from (a) $N=64^2$ to (b) $N=128^2$ and to (c) $N=256^2$, the value of $h$ over which the SW phase dominates over the HC one decreases as $N$ increases, and becomes sharper. In addition, for the larger size (c), a hysteresis phenomenon occurs depending on the history of $h$ (in practice, for clarity, we did not show $p_{\mathrm{SW}}$ in (c)).
}
\label{fig:p0}
\end{figure}

\begin{figure}[tbh]
\includegraphics[]{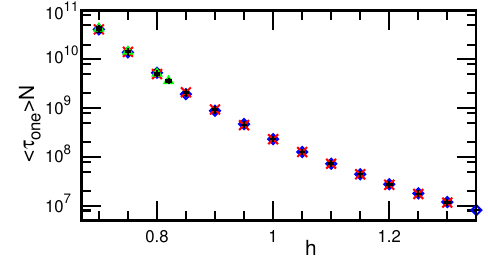}
\caption{
Mean duration $\langle\tau_{\mathrm{one}}\rangle$ (normalized by $1/N$) of homogeneous states filling the whole lattice ($N_s>0.98N$ for $s\in \{0,1,2\}$). 
The symbols $\diamond$, $\times$, and $\triangle$ represents the data at $N=64^2$, $128^2$, and $256^2$, respectively.
}
\label{fig:tfl}
\end{figure}

\begin{figure}[tbh]
\includegraphics[]{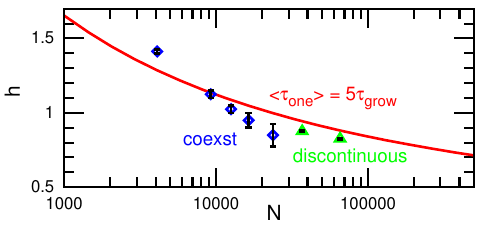}
\caption{
The $h$ threshold separating the HC mode (below the red line) from the SW one (above the red line)  is shown as a function of the system size $N$.  
The squares correspond to the swap of the dominant mode while both modes coexist (see Fig.~\ref{fig:p0}). As $N$ increases, the  triangles are the locus of the discontinuous transition. The solid line represents the condition $\langle \tau_{\mathrm{one}} \rangle = 5\tau_{\mathrm{grow}}$.}
\label{fig:tab}
\end{figure}

\begin{figure}[h]
\includegraphics[width=9cm]{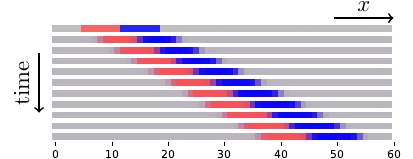}
\caption{
Numerical integration of Eq.~\eqref{eq:mfjb} in one dimension, showing a biphasic $s=2,1$ solitary wave in an $s=0$ gray background. Time flows from top to bottom. Parameters: $\epsilon=4$, $h=0.3$, $\mu=0.46$, $h_a=-0.91$ and $k=1$. Time lapse between two snapshots: $10^5$ time steps $dt=5\times10^{-4}$. The RGB color encodes the system state as $\rho_0\mathcal{G}+\rho_1\mathcal{B}+\rho_2\mathcal{R}$, with $\mathcal{G}$ gray, $\mathcal{B}$ blue, and $\mathcal{R}$ red.
}
\label{fig:jb}
\end{figure}

As the system size is increased, we see that the HC to SW transition evolves from a continuous one ($N=64^2$ and $N=128^2$) to a discontinuous one ($N=256^2$), where it is then accompanied by hysteresis (in practice, this occurs for $N\geq 192^2$). The transition threshold for $h$ decreases with $N$. A rough estimate of the critical value of $h$ seems to indicate a $1/N$ dependence, which would mean that in the $N\to+\infty$ limit, only the SW phase could be observed. For large systems and for $h$ above the threshold, once in the SW phase the system never visits the HC phase, whereas the converse is observed. For small systems (in practice $N \le 154^2$), the SW and HC modes are sequentially visited (see Fig.~\ref{fig:t09}(c)).

Next, we discuss the relative importance of the HC and SW modes. The ratios of HC and SW modes are extracted by measuring the durations of one homogeneous phase and three-phase coexistence, as shown in Fig.~\ref{fig:p0}.
The sum of the one phase and spiral wave ratios is less than unity at $N=64^2$, since the coexistence of two states remains significant.
Note that the fraction of time during which spiral centers occur  ($N_{\mathrm{sp}}>0$) is almost identical to that corresponding to the three phases (the difference is of the order of statistical errors, as shown in Fig.~S1).
{The mean lifetime $\langle\tau_{\mathrm{one}}\rangle$ 
of one homogeneous phase ($N_s>0.98N$)}
decreases with increasing $N$ as $\propto 1/N$  
(see Fig.~\ref{fig:tfl}),
since the nucleation occurs more frequently in larger systems.
The nucleated domain grows to the whole lattice area over a duration $\tau_{\mathrm{grow}}=\sqrt{N}/v_{\mathrm{wave}}$.
The propagation velocity $v_{\mathrm{wave}}$ is roughly linearly dependent on $h$ (see Fig.~S2).
Thus, nucleation of nested domains can occur before  domain growth has been completed over the whole system when $\langle\tau_{\mathrm{one}}\rangle \lesssim \tau_{\mathrm{grow}}$.
This condition accounts for the lower limit of the mode coexistence for small systems  and for a discontinuous transition for large systems (see Fig.~\ref{fig:tab} where the threshold is well fitted by $\langle\tau_{\mathrm{one}}\rangle = 5\tau_{\mathrm{grow}}$). The choice of the Metropolis or Glauber acceptance rate marginally affects the simulation results as shown in Fig.~S3.

\section{Continuum theory}\label{sec:theory}

In order to better identify the minimal ingredients behind the observed phenomenology, we build a mean-field analytical description of the cyclic Potts model, using a grand-canonical free energy and dynamical rules breaking detailed balance. Calling $\rho_i$ the surface density of the state $s=i$, such that $\rho_0+\rho_1+\rho_2=1$, we take the free energy density 
\begin{eqnarray}
 f &=& \sum_{i=0}^2 f_i, \\
f_i &=& \rho_i\ln\rho_i + u_i, \\
 u_i &=& -\mu_i\rho_i-\frac\epsilon2\rho_i^2 +\frac k2\left(\bm\nabla\rho_i\right)^2,
\end{eqnarray}
with $\mu_0=0$ by convention. The first term stems from the entropy of mixing, $\epsilon>0$ describes the attractive coupling between like states and $k>0$ the elastic modulus associated with density gradients. Note that the linear terms can be cast into $-\mu(\rho_1+\rho_2)-\frac12h(\rho_2-\rho_1)$ and interpreted as a chemical potential for the binding particles and an $s=1\to2$ reaction  energy. For well-chosen parameters, this free energy exhibits three minima corresponding to three states $s=i$ of free energies $f_i$ such that $f_1-f_0=f_2-f_1<0$.
For the dynamics, we adopt local equilibrium Glauber rates:
\begin{align}\label{eq:mfjb}
\dot\rho_i=\sum_{j\in\mathcal C_i}\left(
\frac{\rho_j}{1+e^{H_{ji}}}-\frac{\rho_i}{1+e^{H_{ij}}}
\right),
\end{align}
where $\mathcal C_i$ denotes the states complementary to $s=i$, and where the energy variations are given by
\begin{equation}
\begin{split}H_{0i}=\frac{\delta\int d^2r\;u}{\delta\rho_i}=\mu_i-(\epsilon+k\nabla^2)(\rho_i-\rho_0)=-H_{i0},
\end{split}\end{equation}
where $u = \sum_i u_i$.
for the $s=0\to i$ transformation, and $H_{12}=H_{02}-H_{01}=-H_{21}$. Detailed balance is broken by replacing $H_{20}$ by $H'_{20}=H_{20}+h$, thanks to which the cycling step $s=2\to0$ operates even though it is unfavorable regarding the free-energies $f_s$. Note that particle diffusion currents are neglected. The model cycles and produces traveling waves (Fig.~\ref{fig:jb}), which are a one-dimensional proxy to the spiral waves observed in the MC simulations. This confirms that the three competing minima with lifted degeneracy, along with the $h$-driven active mechanism, capture the essential physics of our system. Note that the presence of elastic contributions that penalize interfaces is essential, while direct particle transport is not.

\section{Comparison with predator--prey models}\label{sec:predator}

The formation of spiral waves has been extensively studied using three-species predator--prey simulations.\cite{kerr02,szol14,itoh94,tain94,szab99,szab02,reic07,reic08,szcz13,kels15,mir22}
In the three-species models, each species predates one of the others in the rock--paper--scissors manner.
A crucial difference between our model and predator--prey models is whether sites are updated individually or in pairs.
Two essential processes in the predator--prey models are self-reproduction and predation,
although mobility and other interactions are also considered optionally.
Self-reproduction and predation take place in a pair of neighboring sites as\cite{kerr02,reic07,reic08,szcz13,kels15,mir22}
\begin{eqnarray}
&& A\O \to AA,\ \ B\O \to BB,\ \ C\O \to CC, \\
&& AB \to A\O,\ \ BC \to B\O,\ \ CA \to C\O,
\end{eqnarray}
respectively, where $A$, $B$, and $C$ represent sites occupied by the three species, and $\O$ represents an empty site.
These two processes are often unified as\cite{itoh94,tain94,szab99,szab02}
\begin{eqnarray} \label{eq:pred}
&& AB \to AA,\ \ BC \to BB,\ \ CA \to CC,
\end{eqnarray}
all sites being occupied by one of the species.
In either case, the population increases only by multiplication of one given species.
The flip rate can be a function of the energy difference of two states.\cite{szab02}
However, once one species is extinct, it never appears again.
Although random mutation into other strains can be considered for bacterial strains, the dominant term is still self-reproduction.\cite{szcz13}
Since species stochastically get extinct, the final states are absorption into uniform one-species state.
The flip process in our model is not self-reproduction; in the viewpoint of the predator--prey models, it resembles more directional mutation (flip) with the weight determined by the energy difference.
Thus, extinct species can be raised again and result in the homogeneous cyclic mode.
This recovery is unusual for animal population but is usual for open chemical systems contacted with buffer.
In other words, the single process is accompanied by its reverse process (i.e., it is reversible),
whereas the pair process of the predator--prey model is not (i.e., there is no reverse process from the extinction).
Therefore, thermal-equilibrium-like nucleation and growth 
can be reproduced by the single process but not by the pair process of the predator--prey models.

To confirm  this fundamental difference,
we modified our MC algorithm to replace the flip process by the pair update process~(\ref{eq:pred}), 
the flip rate being still determined by Eq.~(\ref{eq:mc}).
The density of spiral waves (the contacts of three states) becomes lower than that in the single flip process, as shown in Fig.~\ref{fig:nsp}(a),
since the number of domains increases only by domain division in the pair process.
Moreover, the SW mode is obtained only for a finite time period in the pair-process simulations in small systems ($N \leq 128^2$), or low flip energy ($h\lesssim 0.5$ at $N = 256^2$),
so that SW is absorbed into the uniform one-species state eventually.
The mean absorption time is given by $\langle t_{\mathrm{ab}} \rangle= 60~000~000 \pm 10~000~000$ and $120~000~000 \pm 30~000~000$ for $h=1.5$ and $2$, respectively,
at $N = 128^2$ and $\langle t_{\mathrm{ab}} \rangle= 210~000~000 \pm 80~000~000$ at  $h=0.4$  and $N = 256^2$.

The pair update process does not satisfy detailed balance in the thermal equilibrium limit ($h\to0$), unlike the single process.
 Hence, absorption into the uniform one-species state occurs even for conditions yielding a mixing state in thermal equilibrium.
At $h=J=0$, the three states are randomly mixed in thermal equilibrium, which can be reproduced by the single process.
However, the uniform one-species state is obtained by the pair process instead.

\section{Summary}\label{sec:sum}

We have studied the nonequilibrium dynamics of a stochastic cyclic system exhibiting two types of dynamic modes:
a HC mode harboring cyclic changes of homogeneous phases and an SW mode with  three-phase coexistence and spiral waves. The duration of each homogeneous phase is stochastic and, on average, it exponentially decreases with increasing the flipping drive $h$.
The SW dynamics is widely observed in deterministic cyclic systems.
By contrast, the HC dynamics is caused by nucleation and growth, so that it has no counterpart in deterministic models. The transition between the two dynamic modes is achieved by increasing the flipping energy $h$, and we found that it is a continuous crossover in small systems (with coexistence of the two modes), while it becomes discontinuous in large systems.  Nucleation and other stochastic processes can play an important role in small systems (e.g., metal nanoparticles\cite{tang20} and biological systems). In living cells, various types of oscillations and waves have been observed.\cite{beta17,bail22}  We believe that the dynamics we have studied can be experimentally realized in reactions on catalytic surfaces and biological systems. The temporal coexistence of two modes should be observed on the surface of small particles,
whereas the discontinuous transition should occur on macroscopic surfaces. Chemical waves can be accompanied by shape deformation ({\it e.g.} metal particles,\cite{tang20,ghos22} lipid membranes,\cite{wu18,tame21,nogu23a} and gel sheets.\cite{maed08,levi20}) Here, we have considered a non-deformable flat surface. How surface deformations couples to the dynamic modes is a challenging open question.

\begin{acknowledgments}
We thank Hiroyuki Kitahata for informative discussion.
This work was supported by JSPS KAKENHI Grant Number JP21K03481 and JP24K06973. FvW acknowledges the support of the ANR via grant THEMA.
\end{acknowledgments}

\end{document}